\documentclass{iopart}             
\usepackage{iopams}
\usepackage{graphicx}
\usepackage{url,graphics,epsfig}
\usepackage{amssymb}
\usepackage[breaklinks=true,colorlinks=true,linkcolor=blue,urlcolor=blue,citecolor=blue]{hyperref}
\usepackage{cite}

\begin{document}

\jl{2}
%
%
%
\def\etal{{\textit{et al.~}}}
\newcommand{\vdag}{(v)^\dagger}
\newcommand{\myemail}{jbabb@cfa.harvard.edu}
\newcommand{\CHeion}{\textrm{HeC}^+}
\newcommand{\Heatom}{\textrm{He}(1{}^1S)}
\newcommand{\Heion}{\textrm{He}^+({}^2S)}
\newcommand{\Hezero}{\textrm{He}}
\newcommand{\Heplus}{\textrm{He}^+}
\newcommand{\Catom}{\textrm{C}}
\newcommand{\Cion}{\textrm{C}^+}
\newcommand{\CII}{[\Catom\,\textsc{ii}]}
\newcommand{\Oatom}{\textrm{O}}
\newcommand{\Oion}{\textrm{O}^+}
\newcommand{\CatomP}{\textrm{C}({}^3P)}
\newcommand{\OatomP}{\textrm{O}({}^3P)}
\newcommand{\proton}{\textrm{p}}
\newcommand{\phot}{h\nu}
%
%
%
%
%
%
\setlength{\arraycolsep}{2.5pt}             

\title[Radiative charge transfer in collisions of $\Catom$ with $\Heplus$]{Radiative charge transfer in collisions of $\Catom$ with $\Heplus$}

\author{James F Babb$^{1}\footnote[1]{E-mail: jbabb@cfa.harvard.edu}$,
             and B M McLaughlin$^{1,2}\footnote[4]{E-mail: bmclaughlin899@btinternet.com}$}

\address{$^{1}$Institute for Theoretical Atomic and Molecular Physics,
                          Harvard Smithsonian Center for Astrophysics, MS-14,
                          Cambridge, MA 02138, USA}
          
\address{$^{2}$Centre for Theoretical Atomic, Molecular and Optical Physics (CTAMOP),
                          School of Mathematics and Physics, The David Bates Building, 7 College Park,
                          Queen's University Belfast, Belfast BT7 1NN, UK}

%
%
\begin{abstract}
Radiative charge exchange collisions  between a carbon atom $\CatomP$ and a helium ion $\Heion$,
both in their ground state, are investigated theoretically.  Detailed quantum chemistry calculations 
are  carried out  to obtain potential energy curves  and transition dipole matrix elements for 
doublet and quartet molecular states of the HeC$^{+}$ cation.  Radiative charge transfer 
cross sections and rate coefficients are calculated and are found at thermal and lower energies
to be large compared to those for direct charge transfer.
The present results might be applicable to modelling  the complex interplay 
of $\CII$ (or $\Cion$), $\Catom$, and $\textrm{CO}$ at the boundaries of interstellar photon dominated regions (PDRs) and
in xray dominated regions (XDRs), where the abundance of $\Heplus$ 
affects the abundance of $\textrm{CO}$.
\end{abstract}

%
%


\vspace{4cm}

\begin{flushleft}
Short title: Radiative charge transfer in collisions of $\Catom$ with $\Heplus$ \\
\vspace{0.5cm} 
\end{flushleft}


\maketitle
%
%
%
%
\section{Introduction}

Models of carbon monoxide formation in the ejecta of supernova SN 1987a
were found to be sensitive to  the description of  charge transfer collisions
between metal atoms and $\Heplus$,
but  corresponding charge transfer rate coefficients were estimated  due
to the unavailability of  measured or calculated values~\cite{LepDalMcC90,LiuLepDal92}.
To remove this uncertainty, charge transfer cross sections and rate coefficients were calculated for 
$\Catom  + \Heplus$  \cite{KimDalCha93,KimDalCha94} and  
for $\Oatom+\Heplus$ \cite{KimGuLie94,ZhaStaGu04}.
However, according to the results from this early work, the calculated values were found to be too small to
affect appreciably the conclusions reached in the modelling studies.

We note that in the most recent study on $\Oatom+\Heplus$,   Zhao and co-workers \cite{ZhaStaGu04}
found  that  the radiative charge exchange process was more significant compared
to  the nonradiative charge exchange (or direct charge exchange) process.  The dominant channel is
that which leaves the $\Oion$ ion in its ground state, namely,
\begin{equation}
\label{RCT-oxygen}
\OatomP + \Heion \rightarrow \Oion({}^4S^o) + \Heatom + \phot,
\end{equation}
where $\phot$ is the photon energy~\cite{ZhaStaGu04}.
Earlier studies on $\Catom+\Heplus$, by Kimura and co-workers \cite{KimDalCha94}, concluded 
that nonradiative charge exchange was the dominant process 
in comparison to radiative charge exchange which leaves the $\Cion$ in the excited ${}^2D$ state,
\begin{equation}
\CatomP + \Heion \rightarrow \Cion({}^2D) + \Heatom + \phot ,
\end{equation}
though they did not consider the channel---analogous to (\ref{RCT-oxygen})---where
the carbon ion exits in its ground state,
\begin{equation}
\label{RCT-ground}
\CatomP + \Heion \rightarrow \Cion({}^2P^o) + \Heatom + \phot .
\end{equation}
In this paper, we reconsider the radiative charge exchange process for C and $\Heion$,
extending the final states to $\Cion({}^2P^o)$ and $\Cion({}^4P)$.
We find---similarly to $\Oatom+\Heplus$---that  radiative
charge exchange~(\ref{RCT-ground}) 
is considerably larger than direct charge exchange at thermal and lower collisional energies.

The layout of this paper is as follows.  In section 2 we present the theoretical methods used to 
calculate the cross sections and rates.  Section 3 presents the results from our work and provides 
a discussion of our results.  Finally in section 4, conclusions are drawn from our work, 
and we briefly consider implications of these, and other recent results, for metal atoms in 
charge exchange collisions with $\Heplus$  for astrophysical modelling.

%
%
%
%

\section{Theory}

\subsection{Molecular Structure}\label{sec:Theory}
We consider the reaction,
\begin{equation}
\label{rct-rxn}
\CatomP + \Heion \rightarrow \Cion + \Heatom + h\nu,
\end{equation}
representing a collision between a $\CatomP$ atom and a $\Heion$ ion, which  
via an electric dipole radiative transition, results in a residual $\Cion$ ion being  left in one of the final states;  
${}^2P^o$, ${}^4P$, ${}^2D$, ${}^2S$ or ${}^2P$.
Relative to the energy of the initial colliding atom ion pair $\CatomP+\Heion$,
the ${}^2P$ state is slightly above (0.39~eV) the entrance energy,
while the ${}^2P^o$ state lies lower ($-$13.33~eV) compared
to the ${}^2S$ ($-$1.37~eV), ${}^2D$ state  ($-4.04~\textrm{eV}$), and ${}^4P$ state ($-8.00~\textrm{eV}$).
The possible molecular states formed corresponding to the initial  $\CatomP+\Heion$ reactants and the final  
$\Cion + \Heatom$ residual products are listed in Table~\ref{corr-table}.

%
%
%
%
%
\begin{table}
\centering
\caption{The HeC$^+$ cation separated atom ion pair potential energies at an interatomic 
		distance of $R$=12 $\mathrm{a}_0$,  compared to the energies at asymptotically 
	   large internuclear distance from the NIST tabulations \cite{NIST2016}.  
	   The left pointing arrows mark the entrance channels for the atom ion 
        colliding pair.  The Bohr radius $\mathrm{a}_0$ is 
        $5.291\,772\,106\,7 \times$ 10$^{-9}$ cm\label{corr-table}.
        All energies are given in eV. }
\lineup
\begin{tabular}{llcccc}
\br
Separated Atom Ion Pair 		&Molecular States  											& Energy (eV)$^{a}$ 	& Energy (eV)$^{b}$\\
					&												& ($R$=12 $\mathrm{a}_0$) 	& ($R\rightarrow\infty$) \\
\mr
$\Heatom+ \Cion({}^2P)$        	& $\mathrm{4}{}^2\Pi$, $\mathrm{2}{}^2\Sigma^-$    					& 13.64 			& 13.72 \\
$\leftarrow \Heion + \CatomP$	& $\leftarrow\mathrm{2}{}^4\Sigma^-$ 		 					& 13.50			& 13.33 \\
                                                	& $\leftarrow\mathrm{2}{}^4\Pi$ 								& 13.47			& 13.33 \\
                                                	& $\leftarrow\mathrm{3}{}^2\Pi$ 								& 13.24			& 13.33 \\
                                                  	& $\leftarrow\mathrm{D}{}^2\Sigma^-$							& 13.23			& 13.33 \\
$\Heatom+ \Cion({}^2S)$           &$\mathrm{3}{}^2\Sigma^+$                                       					& 11.95 			& 11.96 \\
$\Heatom+ \Cion({}^2D)$          & $\mathrm{C}{}^2\Delta$, $\mathrm{2}{}^2\Sigma^+$,  $\mathrm{B}{}^2\Pi$  	& 9.22 			&  9.29  \\
$\Heatom+ \Cion({}^4P)$           &  $\mathrm{b}{}^4\Pi$    									& 5.33 			&  5.33  \\
                                                	& $\mathrm{a}{}^4\Sigma^-$ 									& 5.32 			&  5.33  \\
$\Heatom + \Cion({}^2P^o)$     & $\mathrm{X}{}^2\Pi$, $\mathrm{A}{}^2\Sigma^+$   					& 0.00			&  0.00  \\
\br	
\end{tabular}
\begin{flushleft}
$^{a}$Present  \textsc{molpro} MRCI+Q calculations. \\
$^{b}$NIST tabulations, Kramida {\it et al.} \cite{NIST2016}.\\
\end{flushleft}
\end{table}

The $\mathrm{X}{}^2\Pi$, $\mathrm{A}{}^2\Sigma^+$, $\mathrm{a}{}^4\Sigma^-$, and $\mathrm{b}{}^4\Pi$ 
potential energy curves (PECs) were calculated by Matoba {\it et al} \cite{MatTanOht08} and by Tuttle  {\it et al} \cite{TutThoVie15}, 
who presented a critical summary of earlier calculations.
Kimura and co-workers \cite{KimDalCha93} calculated the PECs of the  $\mathrm{D}{}^2\Sigma^-$ and $\mathrm{B}{}^2\Pi$ states and
they  extended these calculations to include the PECs of the 
$\mathrm{C}{}^2\Delta$, $\mathrm{3}{}^2\Pi$, $\mathrm{4}{}^2\Pi$, and $\mathrm{2}{}^2\Sigma^-$ states
and the transition dipole moment (TDM) between the $\mathrm{3}{}^2\Pi$  and $\mathrm{B}{}^2\Pi$ states \cite{KimDalCha94}.

Since a complete set of PECs and TDMs  is not available in the literature, for the present study 
we calculated the molecular data for all the states listed in  Table~\ref{corr-table}. 
A brief summary of our structure calculations is provided below.
In our molecular structure work we used a state-averaged-multi-configuration-self-consistent-field (SA-MCSCF) approach, 
followed by multi-reference configuration interaction (MRCI) calculations 
together with the Davidson correction (MRCI+Q)~\cite{HelJorOls00}. 
The  SA-MCSCF method is used as the reference wave function for the MRCI calculations.
We  used  augmented correlation consistent polarised aug-cc-pV6Z (AV6Z) basis sets 
\cite{KenDunHar92,WooDun93,MouWilDun99} in our work as these are known to recover 
approximately  98\% of the electron correlation effects \cite{HelJorOls00} in structure calculations. 
All the PEC and TDM calculations were performed with the quantum chemistry
\textsc{molpro} 2015.1 program package \cite{MOLPRO_brief},  running on parallel architectures.
The PECs and TDM calculations were performed on this system from 1.5~$\mathrm{a}_0$  
to an internuclear bond distance of $12~\mathrm{a}_0$, beyond which they were matched to their 
long range form for dynamical calculations.

%
%
%
%
\begin{figure}
\begin{center}
\includegraphics[width=\textwidth]{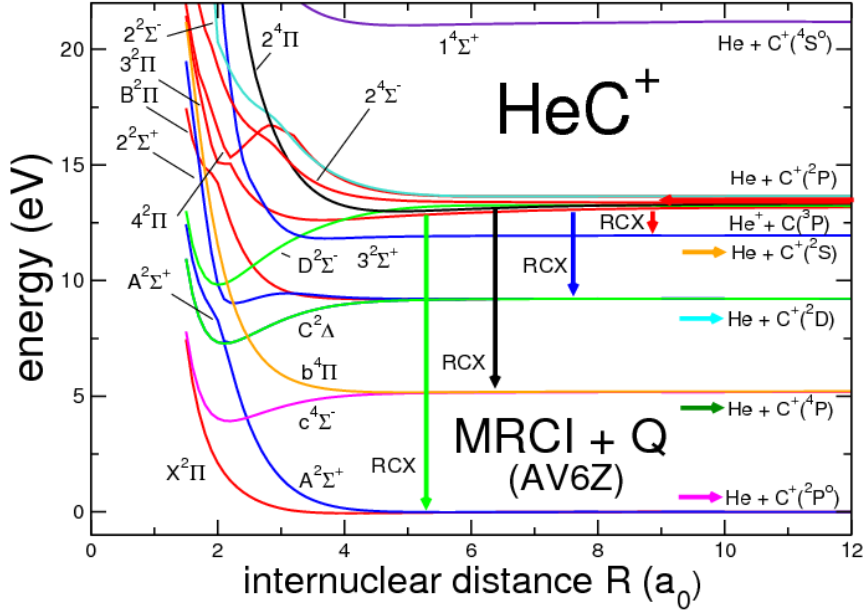}
\caption{Potential energies for the $\CHeion$ molecular ion, 
        as a function of the internuclear distance R (a$_0$),
        corresponding to the initial channel 
        $\CatomP+ \mathrm{He}^+({}^2S) $ and to the 
        final channels of  $\Cion + \mathrm{He}(1{}^1S)$.
        The downward pointing arrows mark the radiative charge exchange (RCX) 
        processes for the dominant doublet and 
        quartet transitions studied here.\label{fig-PEC}}
\end{center}
\end{figure}

%
%
%
%
\begin{figure}
\begin{center}
\includegraphics[width=\textwidth]{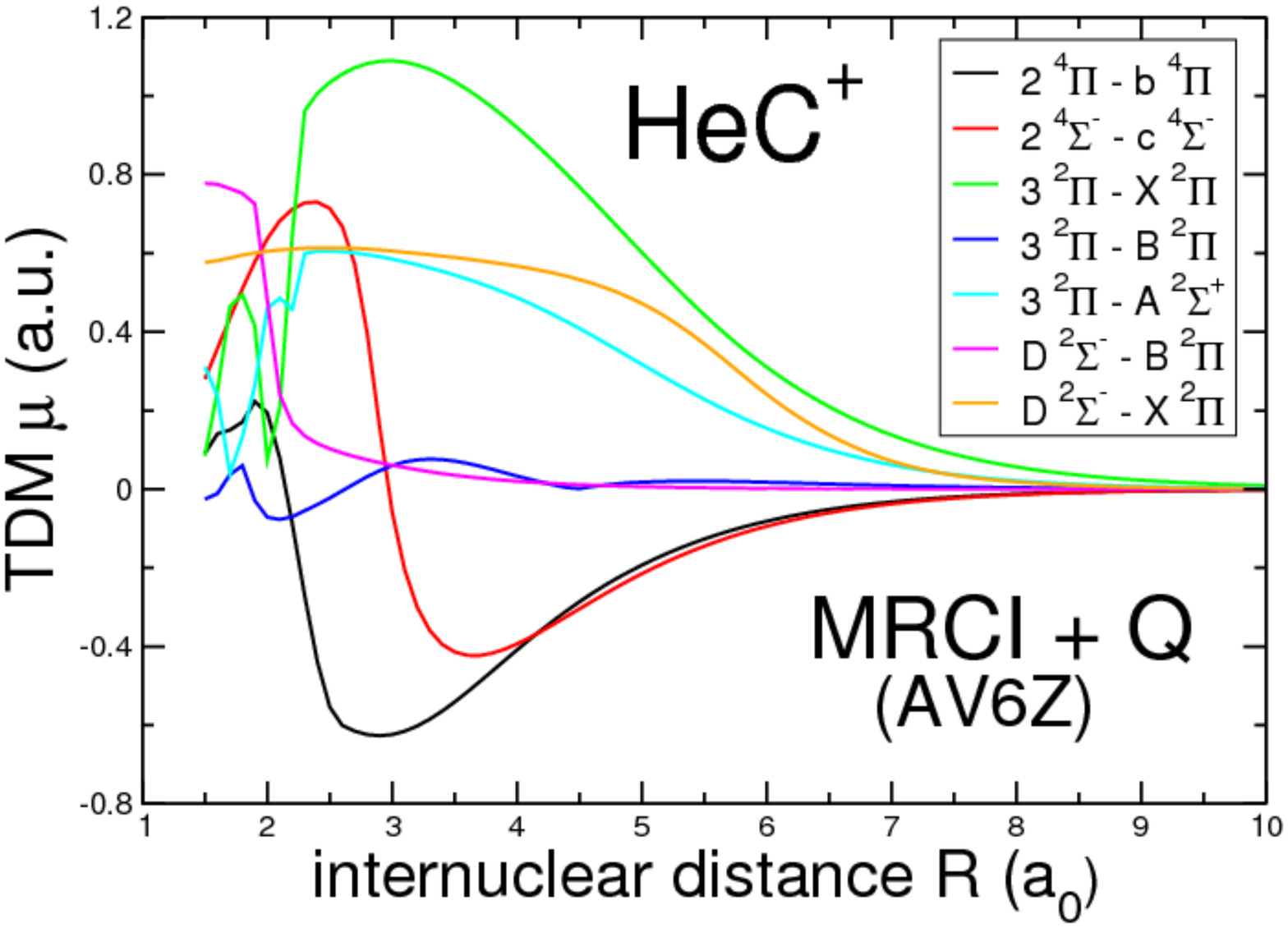}
\caption{Transition dipole moments, in atomic units, 
		 as a function of the internuclear
         distance R (a$_0$), for the $\CHeion$ molecular ion 
         corresponding to the initial channel 
         $\mathrm{C}({}^3P)+ \mathrm{He}^+({}^2S) $ 
         and to the final channel $\mathrm{C}^+({}^2P^o) + \mathrm{He}({}^1S)$. 
         Illustrated are a sample of dominant transitions studied here 
         for the radiative charge exchange processes (RCX).\label{fig-TDM}}
\end{center}
\end{figure}

In  \textsc{molpro} the calculations are carried out in C$_{2v}$ symmetry with the order of Abelian
irreducible representations being ($A_1$, $B_1$, $B_2$,  $A_2$). 
In reducing the symmetry from C$_{{\infty}v}$ to 
C$_{2v}$, the correlating relationships
are $\sigma \rightarrow a_1$, $\pi \rightarrow$ ($b_1$, $b_2$), and 
$\delta \rightarrow$ ($a_1$, $a_2$). 
In order to take account of short-range interactions  
we employed the non-relativistic 
state-averaged complete-active-space-self-consistent-field
 (SA-CASSCF)/MRCI method~\cite{WerKno85,KnoWer85} 
available within the \textsc{molpro} quantum chemistry codes.
In detail, for this cation, six molecular orbitals
(MOs) are put into the active space, including four $a_1$, one $b_1$ and
one $ b_2$ symmetry MO's and the 1$\sigma$ orbital is frozen.
The  molecular orbitals for the MRCI procedure are obtained from the 
state-averaged-multi-configuration-self-consistent-field 
(SA-MCSF)  method,  where the averaging processes for the doublets 
is carried out on the lowest six ($^2A_1$), 
six  ($^2B_1$), and five ($^2A_2$) molecular states of 
this molecule in C$_{2v}$. Separate calculations were 
carried out for the case of the quartets, with the same basis set, where 
the averaging process in this case was performed 
over the lowest two ($^4A_1$), two  ($^4B_1$), 
and two ($^4A_2$) molecular states of this cation.

These MOs ($4a_1$, $1b_1$, $1b_2$, $0a_2$), 
denoted by (4,1,1,0), were  generated from the state-averaging CASSCF process, 
and used to perform all the subsequent PEC calculations for all the 
electronic states in the MRCI+Q approximation.  Fig.~\ref{fig-PEC} 
shows the calculated PECs for the doublet and  quartet states of 
the HeC$^{+}$ cation as a function of bond separation. 
The TDMs considered in the present work are shown in Fig.~\ref{fig-TDM}. A more extensive report on  the calculated  
PECs and TDMs will be presented in a future publication.

For the ground state of the HeC$^{+}$ cation, we find from our calculations 
 that the dissociation energy $D_e$, of the $\mathrm{X}{}^2\Pi$ state is 62~meV, with 
equilibrium bond distance  $r_e$=4.10 $\mathrm{a}_0$.  This is to be compared
to earlier values of $D_e$=50.34 meV and $r_e = 4.24\; \mathrm{a}_0$ obtained by  Grice {\it et al}  \cite{Grice1987}, 
using the \textsc{gaussian} 82 quantum chemistry package at the MP4SDQ/6-311 + G(3$df$,3$pd$) 
level of theory, excluding core contributions to the correlation energy. 
More recent results from higher level approximations give values of $D_e$=58.07 ~meV and of $r_e$=4.18 $\mathrm{a}_0$, 
which were obtained by Matoba {\it et al}  \cite{MatTanOht08}, 
with a diffuse aug-cc-pVQZ  basis (diffuse AVQZ),  within the MCSCF/MRCI approximation using the IBM \textsc{alchemy} II quantum chemistry codes.
A value of $D_e$=59.03~meV and $r_e$=4.16 $\mathrm{a}_0$ was found by Tuttle {\it et al} \cite{TutThoVie15}, 
using the RCCSD(T) approximation,  within the  \textsc{molpro} quantum chemistry package, 
and basis sets of quadruple-$\zeta$ (AVQZ) and quintuple-$\zeta$ (AV5Z) quality; 
with each point counterpoise corrected and extrapolated to the  basis set limit. Our calculations are in good 
agreement with these recent findings.  The value for the dissociation energy $D_e$ is within 
7\% of the value obtained by  Matoba {\it et al}  \cite{MatTanOht08}
and within 5\% of the more sophisticated calculations of Tuttle {\it et al} \cite{TutThoVie15}. This is in quite satisfactory 
agreement  for the present study of collisions at thermal energies between this atom ion pair.

To carry out the dynamical cross section calculations for radiative loss, 
we interpolated the \textit{ab initio} calculated PECs and TDMs 
using cubic splines. For $R <$ 1.5 $\mathrm{a}_0$, 
the \textit{ab initio}  PEC data were connected to 
the analytic form $a\exp(bR)$, where $a$ and
$b$ for each state were determined by fitting. 
For $R >$ 12 $\mathrm{a}_0$, the appropriate long-range forms were used for the separating atom ion pair.
In particular, for $\CatomP + \Heion$, this corresponds to a
$Q_{\mathrm C;\pm} R^{-3}$ quadrupole interaction~\cite{GenGie77}
added to the attractive polarisation potential $-\frac{1}{2}\alpha_{\mathrm C}  R^{-4}$,
where $\alpha_\mathrm{C}$ is the $\CatomP$ atom electric dipole polarisability.
The value of $Q_\mathrm{C}$ is positive for the $\mathrm{3}{}^2\Pi$ state
and negative for the $\mathrm{D}{}^2\Sigma^-$ state,
which can be confirmed by close inspection of Fig.~\ref{fig-PEC}.
The value of $\alpha_\mathrm{C}$ also depends on the state of the C atom.
Both Allison and co-workers~\cite{AllBurRob72} and Miller and Kelly \cite{MilKel72} found that the value of the
polarisability is about 10~\% larger for the $|m_L|=1$ state of $\Catom$,
compared to the $m_L=0$ state, though the values from each publication differ.
To model the $|m_L|=1$ polarisability,
we increased the $m_L=0$ polarisability by ten percent~\cite{AllBurRob72,MilKel72}.
For the $\Cion + \Heatom$ system,
the long-range form is 
$-\frac{1}{2} \alpha_{\mathrm He} R^{-4}$,
where $\alpha_{\mathrm He}$ is the $\Heatom$ polarisability.

In carrying out the dynamical cross-section calculations, we fix the potential energies
at asymptotically large distances to the values listed in Table~\ref{corr-table}.
To expedite continuity of the TDMs for $R< 1.5\;\mathrm{a}_0$
we linearly extrapolated to the intercept at $R=0$.
In practice, this region of the TDMs does not affect the calculations detailed below,
because the PECs are repulsive and the TDMs are relatively small, as
can be seen by inspection of Figs.~\ref{fig-PEC} and \ref{fig-TDM}.
The range of internuclear distances, roughly, between 2 and 6~$\mathrm{a}_0$
is most important for the transition amplitudes, see, for example, Fig.~3 of Ref.~\cite{KimDalCha94}.
For $R> 12\; \mathrm{a}_0$ we fitted the values to the form $R^{-n}$, selecting a value of $n\geq 3$.

\subsection{Dynamics}
We assume the initial channel is $\CatomP$ + $\Heion$. 
Then, there are a number of allowed electric dipole electronic transitions 
originating in the  $\mathrm{3}{}^2\Pi$ or $\mathrm{D}{}^2\Sigma^-$ initial states
to doublet states of lower energy, and similarly for the
initial  $\mathrm{2}{}^4\Sigma^-$ or $\mathrm{2}{}^4\Pi$ states
to quartet states of lower energy.

For $\CHeion$, the $\mathrm{X}{}^2\Pi$ and $\mathrm{A}{}^2\Sigma^+$ states are strongly repulsive
and molecular ion formation by radiative association will be unimportant. 
Here the optical potential theory can be used reliably.
The details of this approximation and its application to various systems are given in several publications
\cite{KimDalCha94,StaZyg96,StaGuHav98,ZhaStaGu04,ZhaWanSta06,LiuQuZho09,LiuQuXia10,ZygLucHud13,McLaughlin2014,SheStaWan15}, 
which the interested reader should consult for further information.

For completeness, we give, in brief, the necessary formulas.
In the optical potential approximation, the cross section for radiative decay from 
a channel with initial state $i$ 
and potential energy $V_i(R)$ to a channel with final state $f$ 
and potential energy $V_f(R)$ is 
\begin{equation}
\label{optical-potential}
\sigma_{fi} (E) = p_i \frac{\pi}{k_i^2}\sum_{J=J_0}^\infty (2J+1)[1-\exp(-4\eta_{fi;J}(E))].
\end{equation}
Here $p_i$ is the probability of approach in the initial state $i$, 
$E=k^2/2\mu$ is the relative kinetic energy, $\mu$ is the reduced mass, 
and $\eta_{fi;J}(E)$ is the imaginary part of the phase shift. 
This phase shift is obtained in the distorted-wave approximation by 
\begin{equation}
\eta_{fi;J}(E) = \frac{\pi\mu}{2k} \int_0^\infty dR\; | s_{i;J}(kR)  |^2 A_{fi}(R),
\end{equation}
where $k= \sqrt{2 \mu [E - V_i (\infty)]}$, and $s_{i;J}(k R)$ is the regular energy normalised
solution of the homogeneous radial equation  \cite{MottMassey1965}.
 The quantity,
\begin{equation}
\label{Einstein}
A_{fi}(R) =(4/3)c^{-3}D_{fi}^2(R) | V_f (R) - V_i(R) |^3 
\end{equation}
is the transition probability.
In Eq.~(\ref{Einstein}), $D_{fi}(R)$ is the TDM between the initial and final
electronic states.
The cross section for  collision-induced radiative decay from the entrance channel, 
i.e., the sum of radiative charge transfer and radiative association, is 
obtained within the optical potential approximation using Eq.~(\ref{optical-potential}).

The rate coefficients $\alpha (T)$, in cm$^3$~s$^{-1}$ as a function of 
temperature $T$ (Kelvin), are obtained by averaging the cross 
section $\sigma_{fi} (E)$ over a Maxwellian velocity distribution and are given by
 \begin{equation}
 \label{thermal-rate}
 \alpha (T) = \left( \frac{8}{\mu \pi} \right)^{1/2} \left( \frac{1}{k_B T} \right)^{3/2} \int_{0}^{\infty} E ~ \sigma_{fi} (E) \exp \left( - \frac{E}{k_B T} \right) d E,
 \end{equation}
where $k_B$ is the Boltzmann constant, $1.380\,648\,52  \times10^{-23}$~J/K.

We note that the relationship $v \sigma_{fi}(E)$ may be used to designate 
an effective energy dependent rate $R(E)$,  in cm$^3$~s$^{-1}$ 
from state $f$ to state $i$, \cite{McLaughlin2014},  where $R(E)$ is given by
\begin{equation}
 R(E) =   \sqrt{2 E/ \mu}  \times {\sigma_{fi}} (E).
 \end{equation}
This form of the energy dependent quasi-rate may be used to estimate  the rate coefficients
by converting $E$ to temperature.

%
%
%

\section{Results and Discussions}
\label{results}
The probability for spontaneous emission, Eq.~(\ref{Einstein}), 
which drives the radiative charge transfer process,
depends on the third power of the photon energy, which for relative
kinetic energies less than several eV is approximately the electronic
potential energy difference between initial and final states.
Thus,  we expect that
the  $\mathrm{3}{}^2\Pi$ to  $\mathrm{X}{}^2\Pi$ and   
$\mathrm{3}{}^2\Pi$  to $\mathrm{A}{}^2\Sigma^+$ 
transitions will be the most important,
as  long as the Franck-Condon
overlap between PECs is favourable and the corresponding TDMs are of order unity (in atomic units).
As we will show, the calculations support this model.
We evaluated Eq.~(\ref{optical-potential}) for the $\mathrm{3}{}^2\Pi$--$\mathrm{X}{}^2\Pi$,
$\mathrm{3}{}^2\Pi$--$\mathrm{A}{}^2\Sigma^+$, 
 $\mathrm{D}{}^2\Sigma^-$--$\mathrm{X}{}^2\Pi$,
$\mathrm{3}{}^2\Pi$--$\mathrm{B}{}^2\Pi$,
and  $2{}^4\Pi$--$b{}^4\Pi$ transitions.
The values of $p_i$ are $\frac{1}{9}$, $\frac{2}{9}$, or $\frac{4}{9}$,
respectively, for the initial 
$\mathrm{D}{}^2\Sigma^-$, $\mathrm{3}{}^2\Sigma^+$ or $2{}^4\Pi$ states.
Detailed results for other possible
transitions, which we expect to be weaker
than the $\mathrm{3}{}^2\Pi$ to  $\mathrm{X}{}^2\Pi$ transition, will be presented in a future publication.

%
%
%
%
\begin{figure}
\begin{center}
\includegraphics[width=\textwidth]{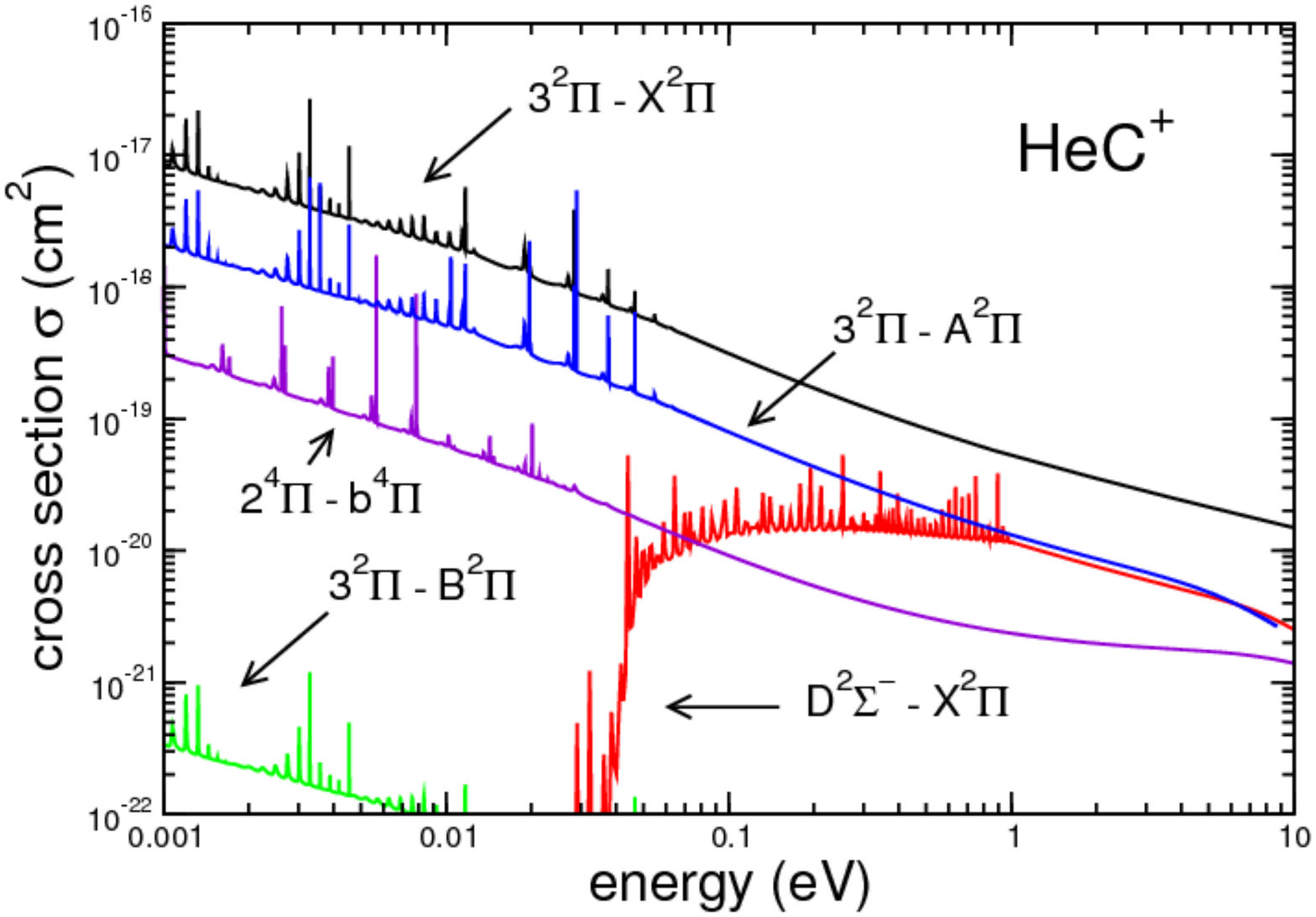}
\caption{Cross sections (cm$^2$) for the radiative loss in collisions from  the initial channel $\CatomP+ \Heion$ 
         to several of the final channels correlating to states of  $\Cion + \Heatom$.
         Ilustrated are the cross sections $ \sigma$ (cm$^2$) as a function of the colliding energy $E $ (eV) 
         for the transitions;  $3{}^2\Pi\rightarrow X{}^2\Pi$ (black line), 
         $3{}^2\Pi\rightarrow A{}^2\Sigma^+$ (blue line), 
         $2{}^4\Pi\rightarrow b{}^4\Pi$ (violet line), 
         and $3{}^2\Pi\rightarrow B{}^2\Pi$ (green line).
         Dropping off in the middle at $E=0.03$~eV, is the 
         $D{}^2\Sigma^-\rightarrow X{}^2\Pi$ transition (red line). \label{cross-fig}}
\end{center}
\end{figure}

%
%
%
%
\begin{figure}
\begin{center}
\includegraphics[width=\textwidth]{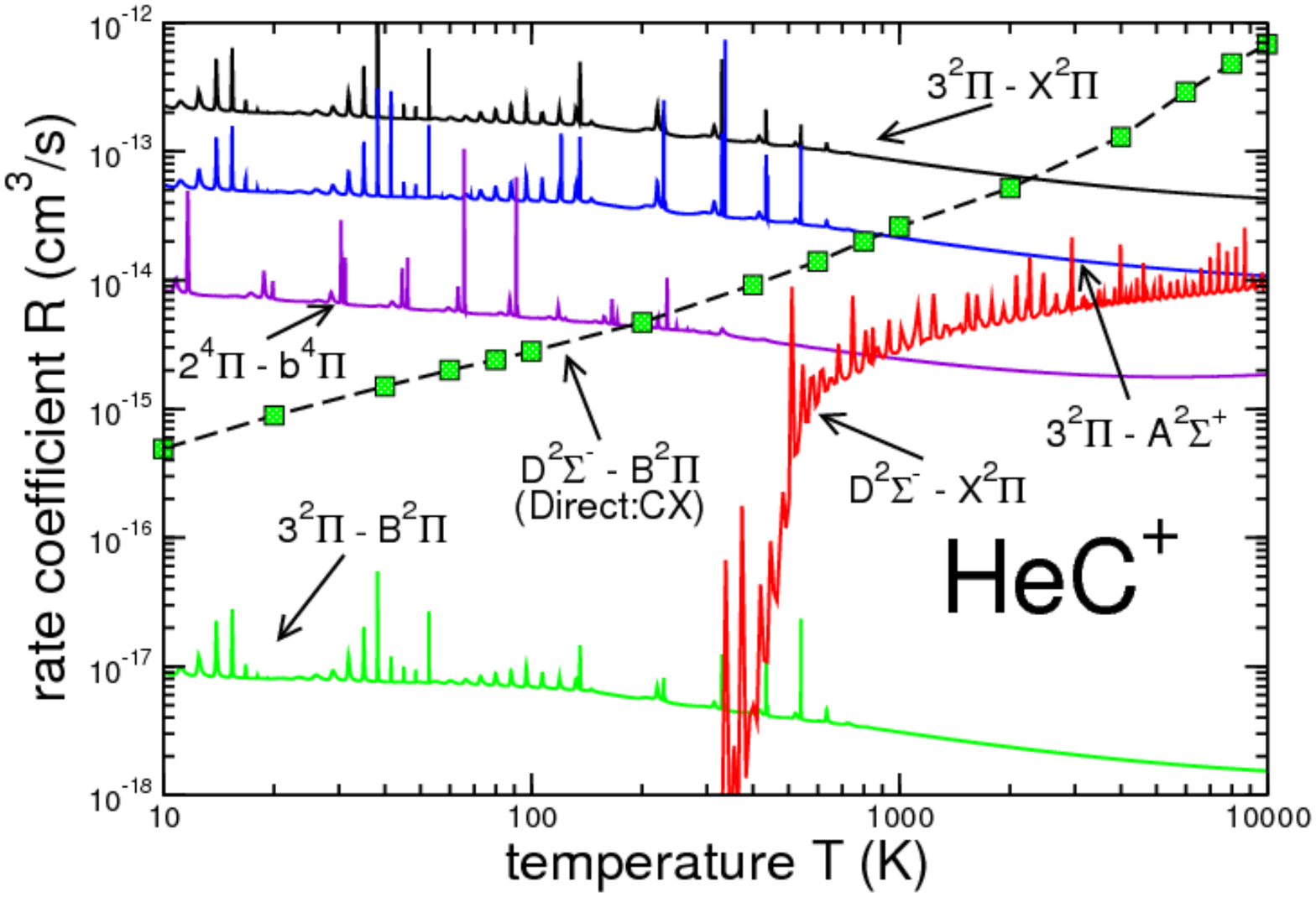}
\caption{Rate coefficients for radiative loss in collisions 
        from  the initial channel $\mathrm{C}({}^3P)+ \Heion$ 
        to several of the the final channels  correlating to 
        states of  $\Cion + \Heatom$.
        Illustrated are the rate coefficients 
        $R(T)$ (cm$^3\mathrm{s}^{-1}$) as a function of temperature $T$ (K),
        for the transitions;  $3{}^2\Pi\rightarrow X{}^2\Pi$ (black line), 
        $3{}^2\Pi\rightarrow A{}^2\Sigma^+$ (blue line),
        $2{}^4\Pi\rightarrow b{}^4\Pi$ (violet line), $3{}^2\Pi\rightarrow B{}^2\Pi$ (green line),  
        and $D{}^2\Sigma^-\rightarrow X{}^2\Pi$ (red line). 
        The dominant $D{}^2\Sigma^-\rightarrow B{}^2\Pi$ 
        channel (dashed black line, with green squares) 
        for direct charge exchange (CX) from the work of 
        Kimura {\textit et al.} \protect\cite{KimDalCha93} is included.}        \label{rate-fig}
        \end{center}
\end{figure}

The calculated cross sections are shown in Fig.~\ref{cross-fig}.
The $\mathrm{3}{}^2\Pi$--$\mathrm{X}{}^2\Pi$ transition is by far the strongest,
as expected, followed by the $\mathrm{3}{}^2\Pi$--$\mathrm{A}{}^2\Sigma^+$ transition.
Numerous resonances,  typical for ion-atom collisions,
occur for collisional energies less than any well depths.
Most of the cross sections follow a power law as the energy $E$ decreases, 
typically, $E^{-1/2}$,
though the $\mathrm{D}{}^2\Sigma^-$--$\mathrm{X}{}^2\Pi$ cross sections
rapidly diminish for energies less than about $0.03$~eV, due to the 
repulsive quadrupole interaction in the  $\mathrm{D}{}^2\Sigma^-$ initial state.
The radiative charge transfer cross sections for the $\mathrm{3}{}^2\Pi$--$\mathrm{B}{}^2\Pi$ transition
are about 100 times larger than those calculated by Kimura {\it et al.}~\cite{KimDalCha94}, but they are still  insignificant.
Direct charge exchange cross sections were calculated by Kimura {\it et al.}~\cite{KimDalCha93}, 
for energies $E>10^{-4}$~eV, and these are seen to vary from about $10^{-21}~\textrm{cm}^2$ at $10^{-4}$~eV
to $10^{-17}~\textrm{cm}^2$ at 10~eV, with the $\mathrm{D}{}^2\Sigma^-$--$\mathrm{B}{}^2\Pi$ channel
the strongest contributor~\cite{KimDalCha94}, driven by rotational coupling~\cite{KimDalCha93}.
We expect that radiative charge transfer via the  $\mathrm{D}{}^2\Sigma^-$--$\mathrm{B}{}^2\Pi$  channel
will diminish with energy similarly
to the $\mathrm{D}{}^2\Sigma^-$--$\mathrm{X}{}^2\Pi$ channel, but because
of the smaller magnitude of the TDM, see Fig.~\ref{fig-TDM},
it will be relatively weak.
Therefore, we conclude for energies less than 
about 1~eV (11\,604.525 Kelvin), radiative 
charge transfer is more significant than direct charge transfer.

%
%
%

The rate coefficients for the $\CatomP + \Heion$ reaction given by 
Eq.~(\ref{rct-rxn})  
are shown in Fig.~\ref{rate-fig}, for the transitions considered here.  
The total rate coefficient for radiative charge transfer 
is about $2\times10^{-13}$ at 10~K, dropping off to about $5\times 10^{-14}$ at $10\,000$~K.
In Fig.~\ref{rate-fig}, we also plot the rate coefficients for direct charge transfer 
from Kimura \textit{at al.}~\cite{KimDalCha93}, which become larger
than the radiative charge transfer process for temperature greater than $3\,000$~K,
approaching $7\times 10^{-13}$ at $10\,000$~K.

\begin{figure}
\centering
\includegraphics[width=\textwidth]{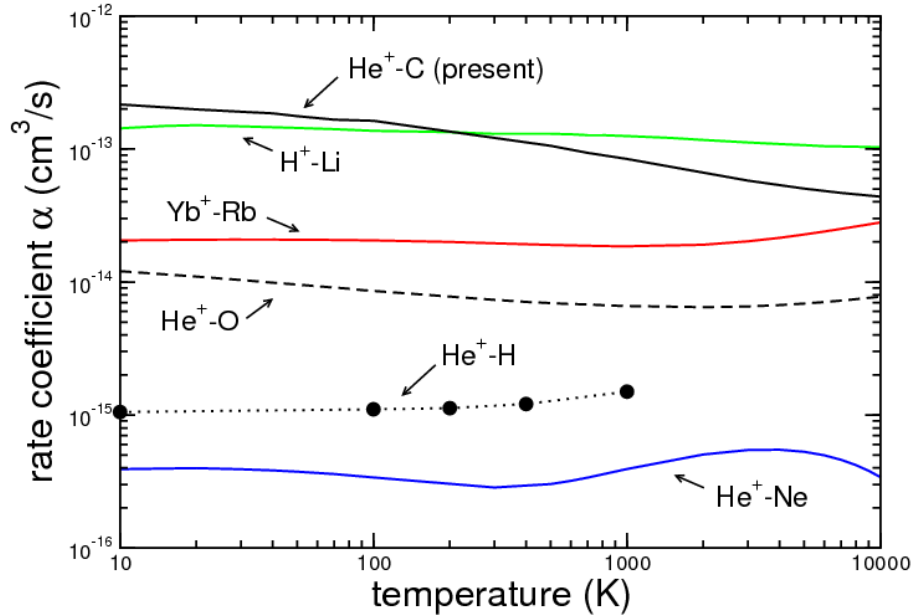}
\caption{A comparison of rate coefficients,  $\alpha(T)$ (cm$^3\mathrm{s}^{-1}$) 
        as a function of temperature $T (K)$, for radiative charge transfer.
        In order of decreasing magnitude at 10~K, 
        $\Heplus$ and $\Catom$, $\mathrm{3}{}^2\Pi$ to  $\mathrm{X}{}^2\Pi$  transition 
         (black line) present work, 
         $\mathrm{H}^+$ and $\mathrm{Li}$ (green line)~\cite{DalKirSta96} ,
         $\mathrm{Yb}^+ +\mathrm{Rb}$ (red line)~\cite{McLaughlin2014},
         $\Heplus$  and $\mathrm{O}$ (dashed black line)~\cite{ZhaStaGu04},
         $\Heplus$  and $\mathrm{H}$~\cite{ZygDalKim89} (points, connected
         with dotted line guide),
         and $\Heplus +\mathrm{Ne}$ (blue line)~\cite{LiuQuXia10}.
         See text for discussion.}
        \label{compare-fig}
\end{figure}

In Fig.~\ref{compare-fig}, we compare
the radiative charge transfer rate coefficient $\alpha(T)$, calculated using Eq.~(\ref{thermal-rate}) for the 
$\mathrm{3}{}^2\Pi$ to  $\mathrm{X}{}^2\Pi$  transition, with the radiative
charge transfer rate coefficients
for $\mathrm{H}^+ + \mathrm{Li}$~\cite{DalKirSta96}, 
$\mathrm{Yb}^+ +\mathrm{Rb}$~\cite{McLaughlin2014},
$\Heplus  +\mathrm{O}$~\cite{ZhaStaGu04},
$\Heplus +\mathrm{H}$~\cite{ZygDalKim89},
and
$\Heplus +\mathrm{Ne}$~\cite{LiuQuXia10}.
The values for $\Heplus +\mathrm{H}$ from Ref.~\cite{ZygDalKim89} were multiplied
by the factor $\frac{1}{4}$ as noted in Ref.~\cite{StaZyg96}.
For $\Heplus +\mathrm{Ne}$ the $\mathrm{B}{}^2\Sigma^+$--$\mathrm{X}{}^2\Sigma^+$
cross sections are more than a factor of 10 larger than the $\mathrm{B}{}^2\Sigma^+$--$\mathrm{A}{}^2\Pi$
cross sections, so the values in the plot were calculated
using the fit from columns 7 and 8 of Table~IV of Ref.~\cite{LiuQuXia10}
for the total radiative charge transfer rate coefficients.
For $\mathrm{H}^+ + \mathrm{Li}$, which has a repulsive ground
state and asymptotic energy difference of about 8~eV between the $2{}^2\Sigma^+$
and $\mathrm{X}{}^2\Sigma^+$states,  the rate coefficients are comparable
to the $\mathrm{3}{}^2\Pi$--$\mathrm{X}{}^2\Pi$
transition of $\mathrm{CHe}^+$ with an asymptotic energy difference of about 13~eV.
The $\mathrm{A}{}^1\Sigma^+$--$\mathrm{X}{}^1\Sigma^+$
transition of $\mathrm{YbRb}^+$ has  a relatively
large TDM~\cite{McLaughlin2014} compared to the $\mathrm{3}{}^2\Pi$ to  $\mathrm{X}{}^2\Pi$  transition
of $\mathrm{CHe}^+$,
but the asymptotic energy difference is only about 2~eV and the reduced mass is 20 times larger.
For $\mathrm{HeH}^+$, the $\mathrm{A}{}^1\Sigma^+$ state is highly repulsive 
and the $\mathrm{X}{}^1\Sigma^+$ state is attractive, leading to comparatively
less favorable transition amplitudes compared to the 
repulsive $\mathrm{3}{}^2\Pi$ and $\mathrm{X}{}^2\Pi$  states of $\mathrm{CHe}^+$.
For $\mathrm{HeNe}^+$, the ground $\mathrm{X}{}^2\Sigma^+$ state has a shallow well
and the asymptotic energy difference for the dominant $\mathrm{B}{}^2\Sigma^+$--$\mathrm{X}{}^2\Sigma^+$
transition is 3~eV, compared to about 13~eV for the $\mathrm{3}{}^2\Pi$--$\mathrm{X}{}^2\Pi$
transition of $\mathrm{CHe}^+$.
The various weight factors (such as we discussed in Sec.~\ref{results}) for each system,
other details of the PECs and TDMS, and the different reduced masses may also contribute
to the relative differences between the rate coefficients.
%

\section{Conclusions}
From the results of our investigations, we find that radiative 
charge exchange (RCX) 
becomes more significant than 
direct charge exchange (CX) as the relative collisional
energy decreases in $\CatomP + \Heplus$ collisions.
Our calculations confirm the earlier finding of Kimura {\it  et al} \cite{KimDalCha94}, 
that radiative charge transfer via the $3{}^2\Pi$--$\mathrm{B}{}^2\Pi$
transition is unimportant.  However, similarly to  work on  
$\Oatom + \Heplus$, by Zhao and co-workers \cite{ZhaStaGu04}, 
our results show that radiative charge transfer
leaving the residual  ion in its ground state is the dominant  mechanism.
Furthermore,  at thermal and lower energies, our results indicate  it is much more rapid
than direct charge transfer.

Earlier calculated rate coefficients for removal of $\Heion$ by $\Catom$
or $\Oatom$ were found to be too small
to affect the ejecta models \cite{KimDalCha93,ZhaStaGu04}.
We note that  charge exchange cross-sections and rates for  collisions of Si with $\Heplus$   were considered recently by 
Satta~\textit{et al.}~\cite{SatGraGia13} using the multi-channel Landau-Zener approximation (MCLZ).
The dominant mechanism is radically different than that for C and O,
due to the presence of a manifold of  excited states $(\mathrm{SiHe}^+)^{\ast}$ above
the exit channel  energy of $\mathrm{Si}^+$ in its ground state.
However, the calculated rate coefficients are not larger than the estimates of the 1990's.
We note that a similar
manifold would be present for the case of charge transfer collisions of S with $\Heplus$, 
but, to our knowledge, the calculation has not been carried out.
Nevertheless, the role  of $\Heplus$ in the destruction of CO is affirmed
by recent ejecta models~\cite{CheDwe09,Cla12}
and it might be interesting to revisit the models of the 1990's
to see if the improved charge exchange rate coefficients
now available for C,  O, or Si with $\Heplus$ 
modify the conclusions obtained at that time.

Furthermore, recent models of ejecta chemistry go beyond  equilibrium chemistry,
but, generally, still  suffer from a lack of charge transfer data~\cite{JerFraKoz11}.
The present results might be applicable to modelling   the complex interplay of $\CII$ (or $\Cion$), $\Catom$, 
and $\textrm{CO}$ at the boundaries of interstellar photon dominated regions (PDRs) and
in xray dominated regions (XDRs), where the abundance of $\Heplus$ can
affect the abundance of CO.

%
%
%

\ack
We would like to dedicate this work to the late 
Professor Alexander Dalgarno, FRS,  who was a great friend, 
a true gentleman and a long term mentor to both authors, 
and from whom we learned a great deal of physics. 
His sharp intellect and foresight into the solution of problems 
will be sadly missed by both the AMO and Astrophysics communities. 
ITAMP is supported in part by a grant from the NSF to the Smithsonian
Astrophysical Observatory and Harvard University. 
B MMcL acknowledges support from the ITAMP visitor's program
and from Queen's University Belfast for the award 
of a Visiting Research Fellowship (VRF).   
Grants of computational time  at the National Energy Research  
Scientific Computing Center (NERSC) in Berkeley, 
CA, USA  and at the High Performance  Computing Center  Stuttgart (HLRS) of the 
University of Stuttgart, Stuttgart, Germany are gratefully acknowledged.
%
%
%

\section*{References}

\providecommand{\newblock}{}

\end {document}